\documentstyle[multicol,prl,aps,epsf,epsfig]{revtex}

\def \b{\beta}

\def \g{\gamma}
\def \d{\delta}

\def \be{\begin{equation}}
\def \ee{\end{equation}}
\def \ben{\begin{eqnarray}}
\def \een{\end{eqnarray}}

\begin{document}

\title{Domination of black hole accretion in brane cosmology}
 
\author{A. S. Majumdar\footnote{Electronic address: archan@bose.res.in}}
 
\address{S. N. Bose National Centre for Basic Sciences,
Block JD, Sector III, Salt Lake, Kolkata 700098, India}
 
\maketitle
 
\begin{abstract}

We consider the evolution of primordial black holes formed during
the high energy phase of  the braneworld scenario.  
We show that the effect of accretion from the surrounding 
radiation bath is dominant compared to evaporation for such black holes. 
This feature lasts till the onset of matter (or black hole) domination
of the total energy density which could occur either in the high energy
phase or later. We find that the black hole evaporation times could
be significantly large even for black holes with small initial mass to
survive till several cosmologically interesting eras. 

PACS number(s): 98.80.Cq
 
\end{abstract}
 
\begin{multicols}{2}

The braneworld scenario in which our observable universe is a brane
embedded in a higher dimensional bulk has gained much popularity in
recent times. The motivation for such a consideration originates from
solutions of string theory where matter and radiation are confined to
the 3-brane, whereas gravity propagates also in the bulk. The significant
consequences of this idea on the physics of the early universe is the
focus of much current attention\cite{langlois}. The foremost departure
from standard cosmology is that there exists a regime during the early
stages when the expansion rate of the universe is proportional to its 
energy density. The aim is to look for observable implications of this 
behaviour on  cosmology.

Primordial black holes through their evolution and evaporation 
are potentially interesting candidates towards probing the high energy
early stages of brane cosmology. The formation of primordial black holes 
through various mechanisms has been studied in much detail\cite{zeldovich}.
More recently, the diverse cosmological ramifications of primordial black 
holes are being vigorously pursued. Several papers have analysed the 
implications of a population of primordial black holes in the early universe
on density perturbations, baryogenesis, reionization and 
nucleosynthesis\cite{barrow,majumdar}. 
Other investigations have focussed on methods of constraining the spectra
of primordial black holes using observational results on  gamma rays, 
background neutrinos, dark matter and quintessence\cite{custodio}. Recently, 
several attempts of describing the formation, evaporation and cosmological 
consesquences of primordial black holes in the braneworld scenario have 
also been undertaken\cite{argyres,guedens}.

In the present Letter, we consider the evolution of primordial black
holes formed during the high energy brane phase when the expansion rate
of the universe is linearly proportional to its energy 
density\cite{langlois}.  We restrict ourselves to the Randall-Sundrum Type
II  model\cite{randall}. Black holes formed during this regime 
evaporate at a rate proportional to their effective area times the 
fourth power of their temperature. Using the induced metric on the brane,
and expressions for the black hole radius and temperature in terms of the
AdS radius $l$, Guedens et al\cite{guedens} derived an evaporation law
which is different from that of usual $4D$ black holes. However, for
a more complete analysis of the evolution of black holes, one must
take into account the effect of accretion from the surrounding radiation
bath as well. This point has been emphasized  by Zeldovich and 
Novikov\cite{zeldovich} and the ramifications
of considering accreting black holes have been found to be significant
in the context of standard cosmology\cite{majumdar}. In our analysis,
we consider the effect of both accretion and evaporation on primordial
black holes formed on the brane.

Let us consider the rate of change of mass $\dot{M}$ of a black
hole immersed in a radiation bath. The accretion rate is proportional to
the surface area of the black hole times the energy density of radiation.
The evaporation rate is proportional to the surface
area times the fourth power of temperature. Thus, $\dot{M}$ is given by
\be
\label{1}
\dot{M} = 4\pi r_{BH}^2\biggl( - g_{brane}\sigma T_{BH}^4 + \rho_R\biggr)
\ee
where $g_{brane}$ represents the effective number of degrees of freedom
on the brane,
$r_{BH}$ and $T_{BH}$ are the radius and temperature of the black
hole, $\sigma$ is the Stefan-Boltzmann constant, and $\rho_R$ is
the energy density of radiation given by
\ben
\label{2}
\rho_R &=& {3M_4^2 \over 32 \pi t_c t} \\
r_{BH} &=& \Biggl({8\over 3\pi}\Biggr)^{1/2}\Biggl({l\over l_4}\Biggr)^{1/2}
\Biggl({M\over M_4}\Biggr)^{1/2}l_4 \\
T_{BH} &=& {1\over 2\pi r_{BH}} \\
\sigma &=& {\Gamma(4)\zeta(4) \over 8(\pi)^2}
\een
where $M_4$ and $l_4$ are the four dimensional Planck mass and length,
respectively, $(t_c \equiv l/2)$ is the transition time to standard cosmology,
and the above relations hold for $t << t_c$.
(Note here that we have neglected the contribution of evaporation into
the bulk given by a term proportional to $4\pi r_{BH}^2g_{bulk}T_{BH}^5$
since it is subdominant even for very small black holes\cite{guedens}).
With the above substitutions the black hole equation can be written as
\be
\label{6}
\dot{M} = -{AM_4^2 \over Mt_c} + {BM \over t}
\ee
where $(A \simeq  3/(16)^3\pi)$ and $(B \simeq 2/\pi)$  are dimensionless 
numbers\footnote{We assume that the black holes can emit massless particles
only and take $g_{brane} = 7.25$ as in  \cite{guedens}.
It is known that  the cross section for
particle absorbtion is increased for relativistic 
particles\cite{zeldovich}, and
also for relativistic motion of black holes\cite{majumdar}. In the
present analysis we neglect these effects which tend to increase the value
 of $B$ by a small amount.}. 
The exact solution for Eq.(6) is 
\be
\label{7}
M(t) = \Biggl[\Biggl(M_0^2 - {2AM_4^2\over 2B -1}{t_0\over t_c}\Biggr)
\Biggl({t\over t_0}\Biggr)^{2B} + {2AM_4^2\over 2B -1}{t\over t_c}\Biggr]^{1/2}
\ee
with $M=M_0$ at the time of BH formation $t=t_0$. It can be seen that
an extremum for the function $M(t)$ exists only if $(M_0/M_4)^2 < (2A/2B-1)(t_0/t_c)$. 
The relation between formation time and mass\cite{guedens}
\be
\label{8}
{t_0\over t_4} \simeq {1\over 4}\Biggl({M_0\over M_4}\Biggr)^{1/2}\Biggl({l\over l_4}\Biggr)^{1/2}
\ee
shows that such a small initial mass will violate the lower
bound on horizon mass $(M_H/M_4) > 2\times 10^6(l_4/l)^{1/3}$ obtained from
the contribution of gravitational waves towards CMBR anisotropy\cite{liddle}.
This result shows that for a radiation dominated high energy phase 
($t < t_c$), accretion
dominates over evaporation for all times and the mass of a 
primordial black hole
continues to grow monotonically with
\be
\label{9}
{M(t) \over M_0} \simeq \Biggl({t\over t_0}\Biggr)^B
\ee

Let us now
consider the scenario in which a certain number density of primordial
black holes exchange energy with the surrounding radiation by accretion
and evaporation. Black holes are formed with an initial mass 
spectrum whose range is much debated\cite{niemeyer}. In order
to keep our analysis uncomplicated, the subsequent calculations are
performed by considering an average initial mass $M_0$. We  assume
that at a time $t_0$ the fraction of the total energy in black holes is
$\b$, and the number density of black holes is $n_{BH}(t_0)$. Hence, we have
\ben
\label{10}
\rho_T(t_0) &=& \rho_R(t_0) + \rho_{BH}(t_0) \\
\rho_{BH}(t_0) = \b \rho_T(t_0) &=& M_0 n_{BH}(t_0) \\
\rho_R(t_0) &=& (1- \b)\rho_T(t_0)
\een

The number density of black holes $n_{BH}(t)$ scales as $a(t)^{-3}$, and 
thus for a radiation dominated evolution on the brane, one has
$(n_{BH}(t)/n_{BH}(t_0)) = (t_0/t)^{3/4}$ since $a(t) \propto t^{1/4}$.
The net energy in black holes grows since accretion dominates over 
evaporation. The condition for
the universe to remain radiation dominated (i.e., $\rho_{BH}(t) < \rho_R(t)$)
at any instant $t$ can be derived from Eqs.(7,10,11,12)
to be
\be
\label{13}
\b < {(t_0/t)^{B+1/4} \over 1 + (t_0/t)^{B+1/4}}
\ee
Thus depending on the value of the parameter $\b$ there may or may not
be an era of black hole domination in the early brane dominated case.
We will analyse these two cases separately.

We first consider the situation when the cosmology stays radiation
dominated up to the time when brane effects are important, i.e., $t\leq t_c$.
For this to be the case, the initial fraction $\b$ should be such that
\be
\label{14}
{\b \over 1 -\b} < \Biggl({t_0 \over t_c}\Biggr)^{B + 1/4}
\ee
If the universe remains radiation dominated up to time $t_c$, then 
(for $M_0 \ge M_4$) the 
mass of a black hole is given by Eq.(9). However, the black hole must
remain small enough, i.e., 
$(M/M_4) < (3\pi/4)(t/t_4)$
for it to obey
the $5D$ evaporation law that we have used\cite{guedens}. These criteria
can be used to put an upper bound on the initial mass:
\be
\label{15}
{M_0\over M_4} < \Biggl({3\pi \over 4(2\sqrt{2})^B}\Biggr)^{{2\over 2-B}}
{t_c \over t_4}
\ee

In the low energy regime $\rho_R \propto t^{-2}$. 
The black hole mass will continue to grow due
to accretion up to a certain time (say $t_t$) after which the radiation
density becomes too dilute for accretion to be significant. At this stage
the rate of evaporation is also insignificant since the black hole masses
have grown by several orders of magnitude from their initial values.
So there ensues
an era during which the black hole mass stays nearly constant over a period
of time until  evaporation takes over finally. This
feature is well supported by numerical simulations of coupled black
hole and Einstein equations for $4D$ black 
holes\cite{majumdar}. For the present case the accretion
rate is smaller since the surface area is $\propto M$ instead of $M^2$
for $4D$ black holes. Furthermore, the
 evaporation rate is ($\propto M^{-1}$) instead of $M^{-2}$. Hence, a 
critical mass $M_{max}$ is reached before evaporation starts dominating
from the transition time $t_t$ onwards\footnote{The exact determination of
$t_t$ requires the numerical integration of the coupled Einstein and black
hole equations (see, for example \cite{majumdar}) and is beyond the
scope of the present paper.}. Assuming the small mass condition
($(M/M_4) < (3\pi/4)(t/t_4)$) to be still valid
together with radiation domination at the time $t_t$, the lifetime for a 
typical black hole can be computed from Eq.(6) (using $M=M_{max}$ at 
$t=t_t$) and is given by
\be
\label{16}
{t_{end} \over t_4} \simeq {1\over 2A}\Biggl({M_{max} \over M_4}\Biggr)^2
{t_c \over t_4}
\ee
Using Eqs.(8) and (9), the lifetime is given by
\be
\label{17}
{t_{end} \over t_4} \simeq {4\over A}(2\sqrt{2})^B\Biggl({M_0\over M_4}\Biggr)^{2-B}
{t_c\over t_4}\Biggl({t_t^2 \over t_c t_4}\Biggr)^B
\ee
Guedens et al\cite{guedens} derived the increase of black hole evaporation 
time  compared to $4D$ black holes as $\propto (l/r_0)^2$, i.e.,
$\{[t_{end}(M_0,5D)]/[t_{end}(M_0,4D)] \simeq [(t_c/t_4)(M_4/M_0)]\}$
Our results show that
 over and above the enhancement of black hole lifetime originating from
a modified evaporation law, the black hole lifetime is further increased
by a factor $\propto (M_{max}/M_0)^2$ purely due to accretion given by
\be
\label{18}
{t_{end}(M,5D) \over t_{end}(M_0,5D)} \simeq \Biggl({M_4\over M_0}\Biggr)^B
\Biggl({t_t^2 \over t_c t_4}\Biggr)^B
\ee
for $t_c \leq t_t \leq t_{eq}$. We thus find that the effect of accretion
in prolonging the life time of black holes is more pronounced compared
to the $5D$ evaporation effect for small black holes. From Eq.(17) it can
be seen that  for 
$(l/l_4) \sim 10^{20}$, a small number density of primordial $5D$ black 
holes formed with very low or even subplanckian masees ($M_0 \leq M_4$) could
survive up to the era of nucleosynthesis and beyond if $t_t \geq 10^5 t_c$. 
This result is likely
to have multifarous cosmological consequences\cite{barrow,custodio}
which need to be investigated in detail. But for black holes with large
initial masses ($M_0 >> M_4$) which form later, there is  
a lower net mass gain. Hence, for example, if we take $t_t \sim 10^5t_c$,
and $(l/l_4) \sim 10^{20}$, black holes formed with 
$M_0 = 10^{15}M_4 \simeq 10^{10}{\rm g}$, will have 
$M_{max} \simeq 10^{15}{\rm g}$,
and will be evaporating now. 
However, if the AdS radius is the maximum allowed by the 
current experimental bounds\cite{hoyle},  i.e., $(l/l_4) \simeq 10^{30}$
  black holes with initial
masses as low as $M_0 = 10^8 M_4 \simeq 10^3{\rm g}$ survive up to the
present era even for $t_t \sim t_c$.

We next consider the case when the initial energy fraction $\b$ is such
that the the accretion process causes $\rho_{BH}(t)$ to exceed $\rho_R(t)$
at a time $t < t_c$. If $\b$ exceeds the bound (we still assume $\b << O(1)$) 
given in Eq.(13), then the 
cosmology enters a matter (black hole) dominated regime in the high energy
phase when  $H^2 \propto \rho^2$. The onset of such a matter dominated era is
labelled by $t_{heq}$ which can be written from Eqs.(9-12) 
as\footnote{Although the formation time $t_0$ of 
individual black holes
is spread out in accordance with the initial mass spectra,
we have used $t_0$ to signify the mean formation time in the similar sense
of using $M_0$ as the mean initial mass.}
\be
\label{19}
{t_{heq}\over t_0} = \Biggl({1-\b \over \b}\Biggr)^{{4\over 4B+1}} \equiv \g
\ee
The mass of a black hole at $t_{heq}$ is given by $M(t_{heq}/M_0) = \g^B$.

For $t > t_{heq}$ the Hubble expansion is essentially driven by the black 
holes ($p=0$ and $H \propto \rho_{BH}$). The number density of black 
holes scales
as matter ($n_{BH}(t) \propto a^{-3}$), 
and thus the scale factor grows as $a \sim t^{1/3}$. During this era, the
radiation density $\rho_R$ is governed by the equation
\be
\label{20} 
{d \over dt}\biggl(\rho_R(t)a^4(t)\biggr) = - \dot{M}(t)n_{BH}(t)a(t)
\ee
where the r.h.s (contribution from accreting black holes)  cannot be
neglected in comparison with the redshiting term ($\rho_R \sim a^{-4}$).
Taking into account the radiation dominated
expansion for  $t < t_{heq}$, and black hole dominated expansion for $t>t_{heq}$,
one has
\be
\label{21}
{n_{BH}(t) \over n_{BH}(t_0)} = \Biggl({t_0\over t_{heq}}\Biggr)^{3/4}{t_{heq}\over t}
\ee
Using Eqs.(9-12) and (21) in (20) gives
\be
\label{22}
\dot{\rho}_R + {4\rho_R \over 3 t} = -\b B \g^{1/4}{\rho(t_0)\over t_0}\Biggl({t
\over t_0}\Biggr)^{B-2}
\ee
Using the condition of matter-radiation equality at $t_{heq}$, i.e.,
$[\rho_R(t_{heq}) = \rho_{BH}(t_{heq}) = M(t_{heq})n_{BH}(t_{heq})]$,
 one can solve Eq.(22) to obtain
\be
\label{23}
\rho_R(t) \approx \g^{-1}\rho(t_0)\Biggl({t_{heq}\over t}\Biggr) - {\b B\over B+1/3}
\g^{1/4}\rho(t_0)\Biggl({t_0\over t}\Biggr)^{1-B}
\ee
where we have neglected a term of higher order in $t_{heq}/t$.

The black hole equation for $t > t_{heq}$ is given by
\be
\label{24}
\dot{M} = B\g^{-1}{M\over t_0}\Biggl({t_{heq}\over t}\Biggr) - C\g^{1/4}{M\over t_0}
\Biggl({t_0\over t}\Biggr)^{1-B}
\ee
where
$C = \b B/(B + 1/3)$
Note that we have ignored the evaporation term at this stage since it is
even more negligible compared to accretion because of the resultant mass 
gain up
to $t_{heq}$. Using $M(t_{heq}/M_0) = \g^B$, the solution of Eq.(24) can
be obtained. 
Beyond $t_{heq}$ the black holes continue to gain mass  until the two terms 
in Eq.(24) become comparable.
At this stage one enters
the transition regime ($t_t$) after which evaporation takes over.
Setting the r.h.s of Eq.(24) to $0$ at $t=t_t$, one obtains
\be
\label{25}
{t_t \over t_{heq}} = \Biggl({B+1/3 \over 1-\b}\Biggr)^{{3\over 3\b +1}} 
\equiv \d
\ee
Thus the accretion regime lasts for a brief duration in the matter dominated
phase, and the maximum  black hole mass $M(t_t)$  is given from the solution
of Eq.(24) by
\ben
\label{26}
{M(t_t) \over M(t_{heq})}  = 
{\rm exp}\Biggl[3B\Biggl(1 - \biggl({\b \over B+1/3}\biggr)^{{1\over 3B+1}}\Biggr) \nonumber \\
+ 
{\b \over B+1/3}\Biggl(1 - \biggl({B+1/3 \over \b}\biggr)^{{3B\over 3B+1}}\Biggr)\Biggr] 
\een

In the evaporating regime ($t>t_t$), the rate of change of black hole mass
is given by the first term in Eq.(6). This can be integrated to give
\be
\label{27}
M(t) = \Biggl[M^2(t_t) - {2AM_4^2\over t_c}(t - t_t)\Biggr]^{1/2}
\ee
The energy density in radiation $\rho_R(t)$ whose time evolution is governed
by Eq.(20) now has a positive contribution from black hole evaporation.
Making the appropriate substitutions, one can again integrate $\dot{\rho}_R(t)$
to obtain
\be
\label{28}
\rho_R(t) \approx 3\b\rho(t_0)\g^{-3/4}{M(t_t)\over M_0}\Biggl[\d^{-1} - {t_{heq}\over t}\Biggr]
\ee
The universe gets reheated as $\rho_R(t)$ increases with time. The stage of
black hole domination lasts up to a time $t_r$ when $\rho_R(t_r) = n_{BH}(t_r)M(t_r)$. 
Subsequently, radiation domination takes over  again. From Eqs.(27) and
(28),
\be
\label{29}
{t_c \over t_r} \approx {3\over 2\d\g} + {A\over 4\g^{2B}}\Biggl({M_4\over M_0}\Biggr)^2
\ee
The standard low energy cosmology  should be radiation
dominated with sufficient radiation temperature for nucleosynthesis to occur.
Thus the era of black hole domination should be over clearly before 
nucleosynthesis. Since it is possible for $t_c/t_4$ to be as large as 
$10^{30}$\cite{hoyle},  let us simply demand that the onset of radiation
domination take place before $t_c$. Requiring $t_r < t_c$, one gets a lower
bound on $\b$ from Eq.(29), i.e.,
\be
\label{30}
\b \geq \Biggl[{4t_0\over 3t_c}\bigl(B + 1/3\bigr)^{{3\over 3B+1}}\Biggr]^{B+1/4}
\ee

The evaporation time of black holes in this scenario can be calculated from
Eq.(27). The mass gain between $t_{heq}$ and $t_t$ given by Eq.(26)
is negligible if Eq.(30) is satisfied and also $\b << O(1)$. The black hole
lifetime is 
\be
\label{31}
{t_{end}\over t_4} \approx \Biggl({M_0\over M_4}\Biggr)^2{t_c\over t_4}\g^{2B}
\ee
The contribution of accretion in prolonging the lifetime of black holes
compared to the modified evaporation effect is given in this scenario by
\be
\label{38}
{t_{end}(M,5D) \over t_{end}(M_0,5D)} =\Biggl({M_{max}\over M_0}\Biggr)^2 \approx \g^{2B}
\ee
This enhancement is independent of the value of the initial mass. Thus
in this case the lifetime of the whole mass spectrum gets prolonged due
to accretion by the same scaling factor $\g^{2B}$, as distinct from the
scenario of  radiation domination throughout where the low mass black holes are
affected most. 
Thus, for instance, if $l/l_4 \simeq 10^{20}$
and $\b = 10^{-3}$, black holes with $M_0 \simeq 10^{12}{\rm g}$ evaporate
during the present era.

To summarize, we have studied the evolution of primordial black holes
formed during the radiation dominated high energy phase on the brane.
These black holes obey a modified evaporation law\cite{guedens}. We have shown
that the accretion of surrounding radiation completely dominates evaporation
as long as radiation domination persists. This results in the net growth
of mass of the black holes. Compared to the case of black holes formed with
the same initial mass in standard cosmology, the black holes in the 
braneworld scenario evaporate much later. We have found that the effect
of accretion in enhancement of black hole lifetime could be very significant
for even low mass black holes for a wide range of parameters. 
It needs to be emphasized that the  effect of accretion and its
consequences of black holes surviving through key epochs in 
cosmology\cite{barrow,custodio} need to be studied in more details. 
Such investigations could  constrain the initial
energy fraction in black holes vis-a-vis the size of the extra dimension.

\end{multicols}
 
\end{document}